\documentclass[letterpaper, 10 pt, conference]{ieeeconf}  

\IEEEoverridecommandlockouts                              
\usepackage{xcolor}                

\usepackage{graphicx} 
\usepackage{blindtext}
\usepackage{dblfloatfix}
\usepackage{graphicx,graphics}
\usepackage{amsmath,amsfonts,mathrsfs,amssymb,dsfont,mathtools}
\usepackage{booktabs}

\usepackage{cite}
\usepackage{bbm}
\usepackage{color}
\usepackage{mathtools}
\usepackage{arydshln}
\usepackage{cuted}
\usepackage{graphicx}

\begin{document}

\title{\LARGE \bf
Analysis of inter-spike interval statistics in neuronal networks with depolarizing and hyperpolarizing threshold potentials
}

\author{Oliver Gambrell$^{1}$ and Abhyudai Singh$^{2}$
\thanks{$^{1}$O. Gambrell is with the Department of Electrical and Computer Engineering, University of Delaware, Newark, DE USA 19716.  \texttt{olg@udel.edu}}%
\thanks{$^{2}$A. Singh is with the Department of Electrical and Computer Engineering, Biomedical Engineering, Mathematical Sciences, Center for Bioinformatics and Computational Biology, University of Delaware, Newark, DE USA 19716. \texttt{absingh@udel.edu}}%
}
\maketitle

\thispagestyle{empty}
\pagestyle{empty}

\begin{abstract}

Neuronal communication is mediated in part by changes in neuronal firing rates.  The time interval between successive neuronal firings is referred to as the inter-spike interval (ISI), and quantifying its statistics is important for understanding neuronal communication.  This paper studies the ISI statistics of a postsynaptic neuron receiving independent excitatory and inhibitory presynaptic action potentials (EI circuit).  This circuit is modeled as a classical integrate-and-fire neuron, and the ISI statistics are investigated for both fixed and adaptive threshold potentials.  First, a depolarizing adaptive threshold model is studied first, where the threshold potential increases with the postsynaptic membrane potential.  Our analysis shows that the ISI noise, quantified as the coefficient of variation, is larger in the adaptive threshold model compared to the fixed threshold model for the same mean ISI.  Additionally, simulations reveal that the ISI noise can be either hypo- or hyper-exponential (defined as ISI noise smaller or larger than one, respectively) depending on the frequencies of excitatory and inhibitory inputs.  Next, a hyperpolarizing adaptive threshold potential is studied, where the threshold decreases as the membrane potential hyperpolarizes. Interestingly, this model shows that the postsynaptic neuron can generate action potentials (APs) when driven solely by inhibitory inputs.  Furthermore, mean and noise signatures are characterized across model parameters for both excitatory and inhibitory inputs.  In summary, this work provides a 
systematic stochastic analysis of adaptive threshold models
for AP generation to understand their role in interneuronal information
processing.

\end{abstract}

\begin{figure*}
\begin{center}
\includegraphics[width=1\linewidth]{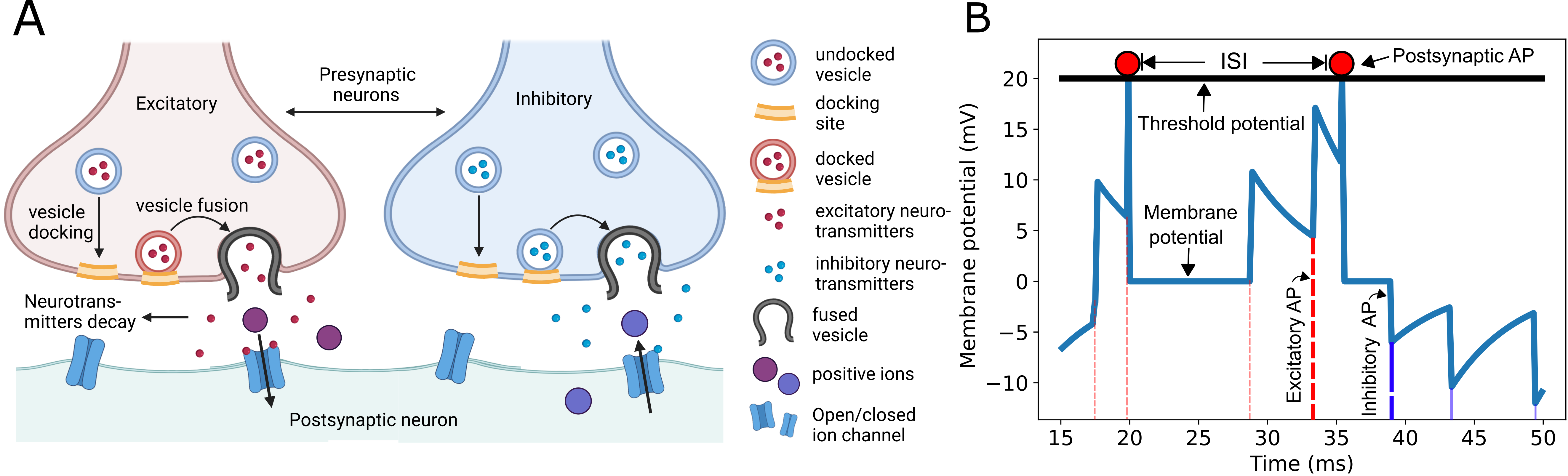}
\end{center}
\caption{\textbf{A schematic of synaptic transmission and a sample trajectory of the membrane potential from the associated synaptic model.}  $\mathbf{A}$:  Synaptic vesicles docked at docking sites at the active zone of the excitatory and inhibitory presynaptic axon terminals fuse with the membrane and release their neurotransmitter content into the synaptic cleft.  These neurotransmitters bind to receptors on the postsynaptic neuron, opening channels that allow ions to flow across the cell and alter its membrane potential. $\mathbf{B}$:  A simulated trajectory of the postsynaptic neuron's membrane potential as modeled in equations \eqref{vplus} and \eqref{dv}.  The membrane potential (solid blue line) depolarizes upon the arrival of excitatory presynaptic action potentials (red dashed line), which arrive via a Poisson process with rate $f_e$.  Inhibitory presynaptic APs (dashed blue line) arrive via a Poisson process with rate $f_i$.  The excitatory and inhibitory QC is modeled as binomial random variables with means $\langle b_e \rangle=\langle b_i \rangle=1$ and noise $CV_{b_e}^2 = CV_{b_i}^2 = 0.95$, respectively.  Between presynaptic APs, the membrane potential returns to a resting potential, here $0$, with time constant $\tau_v$.  Once the membrane potential exceeds a threshold potential $v_{th}$ (black line), the postsynaptic neuron fires an action potential (red circles).  The inter-spike interval (ISI) is the time period between successive postsynaptic APs.  Parameters: $f_e= 100 \ Hz, \ f_i = 100 \ Hz$, $\tau_v=10 \ ms$, $v_{th} = 20 \ mV$, $c_e = c_i = 20 \ mV, \ v_r = 0 \ mV$.}
\label{fig:schemtraj}
\end{figure*}

\section{Introduction}
\noindent{}

The brain is an interconnected network of neurons that communicate through the transmission of electrical signals called action potentials (APs).  As APs arrive at the presynaptic axon terminal, docked synaptic vesicles (SVs) fuse with the membrane and release their neurotransmitter content into the synaptic cleft.  These neurotransmitters diffuse across the synaptic cleft and bind to receptors on ion channels of the postsynaptic neuron, allowing ions to flow across the membrane and alter the membrane potential.  Once the membrane potential crosses a threshold, the postsynaptic neuron fires an AP.  Following this, the membrane potential returns to a resting potential \cite{kandel2021principles}.

The time interval between successive APs, called the inter-spike interval (ISI), is of particular interest in neuroscience, as it reveals information about neuronal firing regularity.  This regularity is primarily controlled by the number of SVs that fuse upon the arrival of a presynaptic AP.  We refer to the number of SVs that fuse as the quantal content (QC).  Previous works have analyzed the statistics of QC including 

The dynamics of SVs and QC and their effect on postsynaptic firing statistics were studied in prior works on auditory synapses \cite{nmul22, brill2021glycinergic, brill2019considerable, vahdat2020information, vahdat2024negative, vahdat22frequency, gambrell2024feedforward, gambrell2025decoy, Pulido2015, Malagon2016}.
Previous works have also shown that steady-state and transient QC follow a binomial distribution under certain limits \cite{rijal2024exact, Bykowska2019, Tsodyks1997, Barri2016, Fuhrmann2002, DeLaRocha2005}.  Furthermore, correlations between successive QC have been shown to be slightly anti-correlated in previous mechanistic models of QC \cite{zahv25}.  Motivated by these observations, we model QC as independent and identically distributed (i.i.d.) random variables.

In previous models of QC, the threshold potential was assumed to be fixed.  This assumption does not account for the probabilistic nature of the threshold potential \cite{Azouz2000} arising from fluctuations in ion channel conductances, such as sodium and potassium \cite{rkoba16}.  To capture this randomness, previous works have used an adaptive threshold 
\cite{Lubejko2019,marasco2023adaptive, chua16,  LEVY2020534}.
In this paper, our goal is to model the random nature of the threshold potential by allowing it to change in response to presynaptic activity.  Using a classical integrate and fire model, we consider an excitatory and an inhibitory presynaptic neuron independently firing AP onto a postsynaptic neuron under the assumption of a fixed threshold potential.  We refer to this model as an EI model.  The firing behavior of EI models has been studied in prior works on excitatory-inhibitory balance \cite{CROCCO2025102646, Alizadeh2025layer, CohenKadosh2025EI, PhysRevLett.134.068403, Chakravarty2025EI}, within neuronal assemblies \cite{ssadeh21, Rich2020neuromodulation, Dzyubenko2021ECM}, and with stochastic models \cite{Reyes2026}.

In this paper, a first-passage-time framework \cite{gds17, aa16, gvs15} is used to derive an equation for the time period between two successive postsynaptic APs, known as the inter-spike interval (ISI).  FPT problems have been studied in detail in biological systems including gene expression \cite{Co2017stochastic, Biswas2021fpt, Ham2024timing, Ali2022architecture, Gupta2018precision, Biswas2024, dey21, zvah21}, lysis timing \cite{Kannoly2025,SKannoly22, Singh2014holin, Mondal2024lysis},  cellular event timing \cite{Ghusinga2016division, Ghusinga2017bursts}, and in prior works on neuronal modeling \cite{ogam25}.  Our FPT analysis show that the noise in the ISI, defined as the coefficient of variation squared of the ISI, is maximized when the excitatory and inhibitory input frequency are both high and approximately equal and the membrane potential remains fixed between presynaptic APs. Next, we allow the threshold potential to increase in response to presynaptic excitation \cite{rw05}, 
which corresponds biologically to the phenomena where a neuron's firing frequency decreases after prolonged stimulation \cite{Liu2001}.  This effect is partly caused by the dynamics of sodium and calcium-activated potassium channels \cite{Orfali2023, Ha2017, rkoba16}.  We refer to this model as a depolarizing adaptive threshold. 

Our analysis shows that the ISI noise, which we model with the coefficient of variation, is larger for a neuron with a depolarizing adaptive threshold than a neuron with a fixed threshold when the mean ISIs of the two models are equal.  Furthermore, simulations reveal that different combinations of excitatory and inhibitory input frequencies can result in the ISI noise being either hypo-exponential (ISI noise is less than one) or hyper-exponential (ISI noise is greater than one).

Next, we allow the threshold potential to decrease in response to presynaptic inhibition, modeling the phenomena of postinhibitory rebound \cite{MAHROUS20251906, Schmid2024}.  The lowering of the threshold potential corresponds biologically to sodium channel recovery from inactivation \cite{Bean2007}.  We refer to this model as a hyperpolarizing adaptive threshold.  Interestingly, this model shows that inhibitory presynaptic APs can elicit postsynaptic APs in the absence of presynaptic excitation.  We now formulate the models.  A summary of the model parameters is in  Table~\ref{Table:parm}.

\begin{table}[t!] 
 	\caption{Summary of model parameters}
	\centering
 \setlength{\tabcolsep}{3pt}
\begin{tabular}{|p{40pt}|p{30pt}|p{145pt}|}
		\hline neuron  & 
	parameter & \hspace{50pt} description \\
		\hline &\hspace{8pt}
 $b_e$ & number of SVs which fuse at the excitatory presynaptic axon terminal\\ 
      & \hspace{7pt} $b_i$ & number of SVs which fuse at the inhibitory presynaptic axon terminal

\\presynaptic  & \hspace{8pt} ${f_e}$ & the rate that action potentials arrive at the excitatory presynaptic axon terminal , $Hz$  \\ 
 & $ \hspace{10pt} {f_i}$ & the rate that action potentials arrive at the inhibitory presynaptic axon terminal , $Hz$ 
\\ \hline & \hspace{8pt} $v$ &  membrane potential, $volts$\\ & \hspace{7pt} $\tau_v$ & membrane time constant, $sec$ \\ & \hspace{8pt} $v_r$ & resting potential, $volts$\\ & 
\hspace{8pt} $v_{th}$ & threshold potential, $volts$\\ & \hspace{8pt} $T$ & inter-spike interval (ISI), $sec$\\ postsynaptic& 
\hspace{8pt} $CV_T^2$ & ISI noise, coefficient of variation squared\\ & \hspace{8pt} $c_e$  & excitatory quantal size, $volts/\text{vesicle}$ \\ &
\hspace{8pt} $c_i$ & inhibitory quantal size, $volts/\text{vesicle}$ \\ & 
\hspace{8pt} $\tau_{th}$ & Threshold time constant, $sec$ \\ & 
\hspace{5pt} $v_{th,\infty}$ & steady state threshold potential, $volts$\\ &
\hspace{8pt} $v_{b}$ & base threshold potential, $volts$ \\ & 
\hspace{8pt} $a_T$ & scaling parameter \\ & 
\hspace{8pt} $v_I$ & inflection potential, $volts$ \\ & 
 \hspace{8pt} $v_{low}$ & minimium threshold potential,  $volts$\\ &
 \hspace{8pt} $v_1$ & limiting potential, $volts$ \\ 
\hline 	\end{tabular}\label{Table:parm}
\vspace*{-4mm}
\end{table}
\section{Model formulation of an excitatory-inhibitory neuronal network}

A schematic of the system to be modeled is presented in Fig.~\ref{fig:schemtraj}-A.  The model, which we refer to as the excitatory-inhibitory (EI) model, was analyzed in \cite{gambrell2026adaptive} and will be restated here for convenience.  The arrival times of excitatory and inhibitory presynaptic action potentials (APs) are modeled as Poisson processes with rates $f_e$ and $f_i$, respectively.  Upon the arrival of an excitatory or inhibitory presynaptic AP, synaptic vesicles (SVs) docked at the active zone of the associated presynaptic neuron fuse with the membrane.  The number of docked SVs that fuse at the excitatory and inhibitory axon terminals is denoted as $b_e$ and $b_i$, respectively.  The random variables $b_e$ and $b_i$ are referred to as the excitatory and inhibitory quantal content (QC).    For modeling simplicity, $b_e$ and $b_i$ are assumed to be independent and identically distributed (i.i.d.) binomially distributed random variables.  

Upon the fusion of SVs, neurotransmitters are released into the synaptic cleft and bind to receptors on postsynaptic membrane.  This binding initiates conformational changes in ion channels, allowing ions to flow across the membrane and change the neuronal membrane potential.  Representing the membrane potential with a continuous random process $v(t)$, the change in $v(t)$ due to excitatory and inhibitory QC $b_e$ and $b_i$ is modeled with the following stochastic resets:
\begin{equation}
    v(t) \xrightarrow[]{f_e}v(t) + c_e \times b_e, \quad v(t) \xrightarrow[]{f_i} v(t) - c_i \times b_i. \label{vplus}
\end{equation}
\noindent In equation \eqref{vplus}, the parameters $c_e$ and $c_i$, called the excitatory and inhibitory quantal size, represent the average amount of depolarization and polarization the membrane potential experiences per SV fusion, respectively. 

Between presynaptic APs, $v(t)$ returns to a baseline resting potential $v_r$ via the classic integrate-and-fire model:
\begin{equation}
    \frac{dv(t)}{dt} = -\left(\frac{v(t)-v_r}{\tau_v}\right), \label{dv}
\end{equation}
\begin{figure*}
\begin{center}
\includegraphics[width=0.8\linewidth]{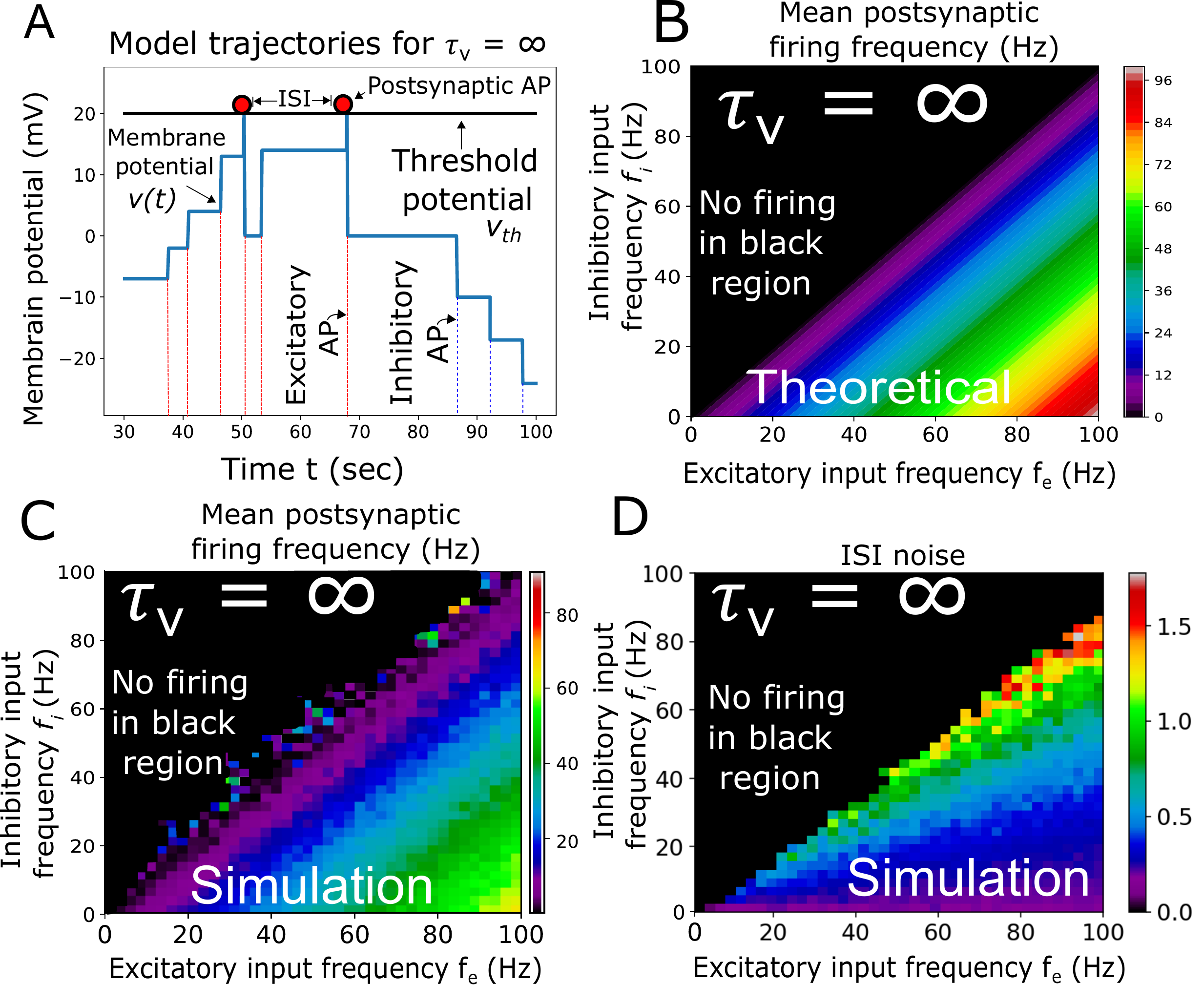}
\end{center}
\caption{\textbf{
The mean postsynaptic firing frequency and inter-spike interval (ISI) noise profile vary with the membrane time constant.
} $\mathbf{A}$:  Trajectories of the membrane and threshold potential in \eqref{vplus} and \eqref{dv} as a function of time when $\tau_v = \infty$.  The membrane potential (blue line) increases if the arriving presynaptic AP is excitatory and decreases if the arriving presynaptic AP is inhibitory.  Upon being driven to the threshold potential $v_{th}$ (black line), the postsynaptic neuron fires an AP (red dot) and resets to a resting potential $v_r = 0 \ mV$.  $\mathbf{B}$:  Theoretical predictions of the mean postsynaptic firing frequency as a function of the excitatory and inhibitory input frequency based on equation \eqref{ttauinf}.  $\mathbf{C}$:  Simulations of the mean postsynaptic firing frequency as a function of the excitatory and inhibitory input frequency in \eqref{vplus} and \eqref{dv}.    $\mathbf{D}$:  Simulations of the ISI noise for a pure integrator as a function of the excitatory and inhibitory input frequency.  The excitatory and inhibitory QC for each of these plots is modeled as binomial distributions with means $\langle b_e \rangle  =1$ and noise $CV_{b_e}^2 = 0.95$.   Parameters: $\tau_v=100 \ ms$, $\tau_{th}=100 \ ms$, $v_b = 20 \ mV$, $v_{min} = -50 \ mV$, $v_I = 10 \ mV$, $a_T = 0.8 \ mV$, $c_e = c_i= 2 \ mV, \ v_r = 0 \ mV$.}
\label{fig:ft}
\end{figure*}
where $\tau_v$ is the positively valued membrane time constant that controls the rate at which the membrane potential returns to $v_r$.  For modeling simplicity, we assume that $v_r=0$.  Once $v(t)$ crosses a threshold potential $v_{th}$, the postsynaptic neuron fires an AP and is assumed to instantly reset to $v_r$.  In this section the threshold potential is set to a fixed value, and in later sections $v_{th}$ will be modeled as a random process. Trajectories of the random process $v(t)$, which is subject to the stochastic resets in \eqref{vplus} and the dynamics in \eqref{dv}, are plotted in Fig.~\ref{fig:schemtraj}-B.  Having defined the EI model, we now quantify postsynaptic statistics of interest.

The time interval between two successive postsynaptic APs is called the inter-spike interval (ISI), and is defined~as 
\begin{equation}
    T := \inf\{t\ge 0 : v(t)\ge v_{th} \text{ and } v(0)=v_r\}.\label{tmean}
\end{equation}
Both the mean ISI, $\langle T \rangle$, and the ISI noise, defined as
\begin{equation}
    CV_T^2 := \frac{\langle T^2 \rangle - \langle T \rangle^2}{\langle T \rangle^2}, \label{cv2}
\end{equation}
are analyzed, as studying these metrics together provides insight into the mechanisms of neuronal processing.  In equation \eqref{cv2}, the notation $\langle \ \rangle$ denotes the expected value operator.  The problem of determining the statistics of $T$ is referred to as a first-passage-time (FPT) problem, and FPT problems have been analyzed in many biological systems \cite{gds17, SKannoly22, Co2017stochastic, Ghusinga2016division}, including prior works on neuronal modeling \cite{gambrell2025decoy}. Having defined the model, we now analyze the statistics of its ISI. In particular, we study the ISI noise and the inverse of the mean ISI, called the mean postsynaptic firing frequency. The mean postsynaptic firing frequency is studied instead of the mean ISI for ease of analysis.

\section{Mean postsynaptic firing time analysis of the excitatory-inhibitory neuronal model}

In the previous section, we formulated the model of an excitatory and an inhibitory presynaptic neuron independently firing APs onto a postsynaptic neuron, which we called the EI model.  The ISI statistics for the EI model are now analyzed.  We begin by first defining a few terms:
\begin{itemize}
    \item $N(t) \ - $ The number of presynaptic APs which have occurred up to time $t$.
    \item $u_j \ - $ The change in $v(t)$ due to the $j$-th presynaptic AP.  $u_{j}=c_e b_e$ if the $j$-th presynaptic AP is excitatory, and $u_j=-c_i b_i$ if inhibitory.
    \item $t_j \ -$ The arrival time of the $j$-th presynaptic AP.
    \item $t_{N(t)} \ -$ The arrival time of the $N(t)$-th presynaptic AP.
\end{itemize}

Using these definitions and equations \eqref{vplus} and \eqref{dv}, we can write $v(t)$ as 
\begin{equation}
    v(t) = \sum_{j=1}^{N(t)}u_je^{-\frac{t_{N(t)}-t_{j}}{\tau_v}}.\label{veq}
\end{equation}
For simplicity, we solve for $\langle T \rangle$ for the specific case of a classical pure integrator $(\tau_v \to \infty)$ (Fig.~\ref{fig:ft}-A) and assume a small noise approximation.

When $\tau_v\to\infty$, equation \eqref{veq} becomes
\begin{equation}
    v(t) = \sum_{j=1}^{N(t)}u_j.\label{vinftau}
\end{equation}
Taking the expected value of \eqref{vinftau} yields
\begin{equation}
    \langle v(t) \rangle = \left \langle \sum_{j=1}^{N(t)}u_j\right \rangle = \left \langle \left \langle \sum_{j=1}^{N(t)}u_j \middle|N(t)\right \rangle \right \rangle = \langle N(t) \rangle \langle u_j \rangle. \label{vmean}
\end{equation}
\begin{figure*}
\begin{center}
\includegraphics[width=0.7\linewidth]{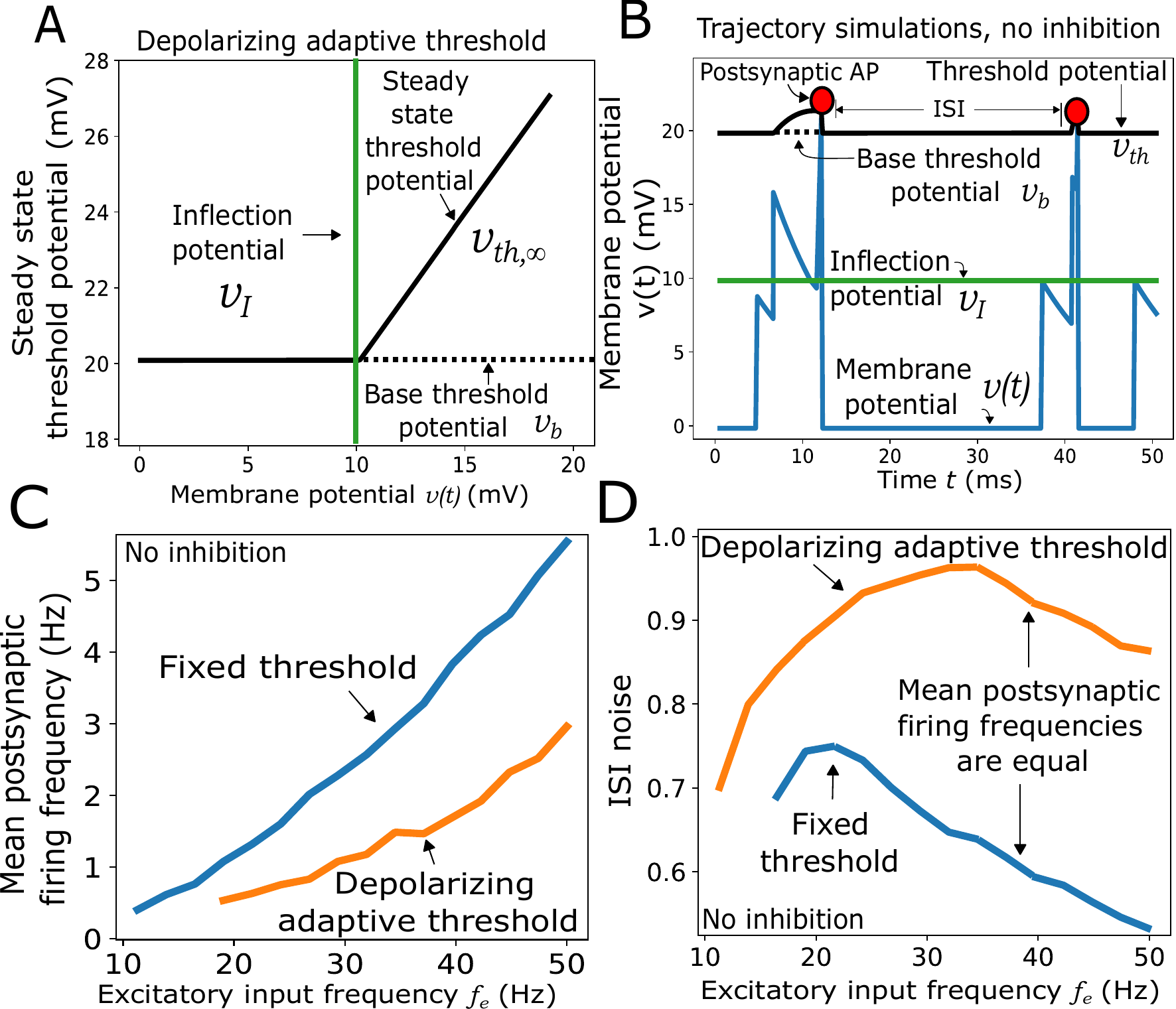}
\end{center}
\caption{\textbf{The inter-spike interval (ISI) noise is larger in neurons with a depolarizing adaptive threshold than those with a fixed threshold.}  $\mathbf{A}$:  The steady state threshold potential as a function of the membrane potential as described in equation \eqref{vthpos}.  $\mathbf{B}$: Trajectories of the depolarizing adaptive threshold and membrane potential as a function of time.  The membrane potential (blue line) evolves via \eqref{vplus} and \eqref{dv} while the threshold potential (black line) evolves via \eqref{dvth1} and \eqref{vthpos}.  In these simulations, the postsynaptic neuron only receives excitation $(f_i=0)$.  Once the membrane potential is driven to the threshold potential, the postsynaptic neuron fires an AP (red dot).  The inter-spike interval (ISI) is the time interval between successive postsynaptic APs. $\mathbf{C}$:  A comparison of the mean postsynaptic firing frequency as a function of the excitatory input frequency for both a fixed threshold and the depolarizing adaptive threshold.  $\mathbf{D}$:  The ISI noise as a function of the excitatory input frequency for both a fixed threshold and a depolarizing adaptive threshold model.  For each value of $f_e$, the mean postsynaptic firing frequencies of the two models were matched by incrementally adjusting the adaptive threshold parameter $v_b$ until the difference between the mean postsynaptic firing frequencies was within a $10 \ \mu Hz$ difference.  The excitatory and inhibitory QC for each of these plots is modeled as binomial distributions with means $\langle b_e \rangle  =1$ and noise $CV_{b_e}^2 = 0.95$.   Parameters: $\tau_v=100 \ ms$, $\tau_{th}=100 \ ms$, $v_b = 20 \ mV$, $v_{min} = -50 \ mV$, $v_I = 10 \ mV$, $a_T = 0.8 \ mV, \ c_e = 2 \ mV, \ v_r = 0 \ mV$.}
\label{fig:dat}
\end{figure*}
We can solve for $\langle T \rangle$ by observing that the arrival times of presynaptic APs are modeled via Poisson processes, implying that the probability of an excitatory presynaptic AP is $\frac{f_e}{f_e+f_i}$ and $\frac{f_i}{f_e+f_i}$ for an inhibitory presynaptic AP.  Using this and the fact that $N(t)$ is the sum of two Poisson processes, we can show that 
\begin{equation}
    \langle v(t) \rangle = t\left(c_e f_e \langle b_e \rangle-c_i f_i \langle b_i \rangle \right).\label{vinf}
\end{equation}
Making the approximation that $\langle v(\langle T \rangle)\rangle \approx v_{th}$, which is valid when the noise in $v(t)$ is small, we have
\begin{equation}
    \langle T \rangle \approx \frac{v_{th}}{c_e f_e\langle b_e \rangle-c_i f_i\langle b_i \rangle },\label{ttauinf}
\end{equation}
which is defined for $c_e f_e\langle b_e \rangle >c_i f_i \langle b_i \rangle$.  

Fig.~\ref{fig:ft}-B and Fig.~\ref{fig:ft}-C show the mean postsynaptic firing frequency for the theoretical equation in \eqref{ttauinf} and simulations of the model in equations \eqref{vplus} and \eqref{dv} for $\tau_v\to\infty$, respectively.  These plots show that inhibition always reduces the mean postsynaptic firing frequency.  Furthermore, when $f_i>f_e$, both simulations and theoretical predictions show that postsynaptic APs do not occur on average.  We resort to simulations for noise in Fig.~\ref{fig:ft}-D, which shows that noise is maximized when $f_e$ and $f_i$ are both large and $f_e>f_i$.

When $\tau_v\to 0$, the membrane potential instantly returns to its resting potential after the postsynaptic neuron fires an AP, implying that inhibition has no effect on the ISI statistics.  The mean ISI and ISI noise for the case of $\tau_v\to0$ were solved for previously in \cite{Gambrell2025}.  The solution was derived by representing the ISI as the sum of the times between successive presynaptic APs, and then computing its moments.  We present the results here for simplicity:
\begin{equation}
    \langle T \rangle = \left[f_e\times\left( 1-\mathbf{P}\left(b_e\le \frac{v_{th}}{c_e}\right)\right)\right]^{-1},
\end{equation}
\begin{equation}
    CV_T^2=1,
\end{equation}
where $\mathbf{P}$ is the probability mass function of $b_e$.  The case of $0<\tau_v< \infty$ has been explored in \cite{gambrell2026adaptive}, which showed that the mean postsynaptic firing frequency is maximized when $f_i$ is small and $f_e$ is high.  Additionally, ISI noise was shown to be maximized at intermediate excitatory input frequencies.

The model in this section assumed a fixed threshold potential.  We now focus on the case of pure excitation and allow the threshold potential to increase in response to postsynaptic activity.  We refer to this type of threshold as a depolarizing adaptive threshold.

\begin{figure*}
\begin{center}
\includegraphics[width=0.95\linewidth]{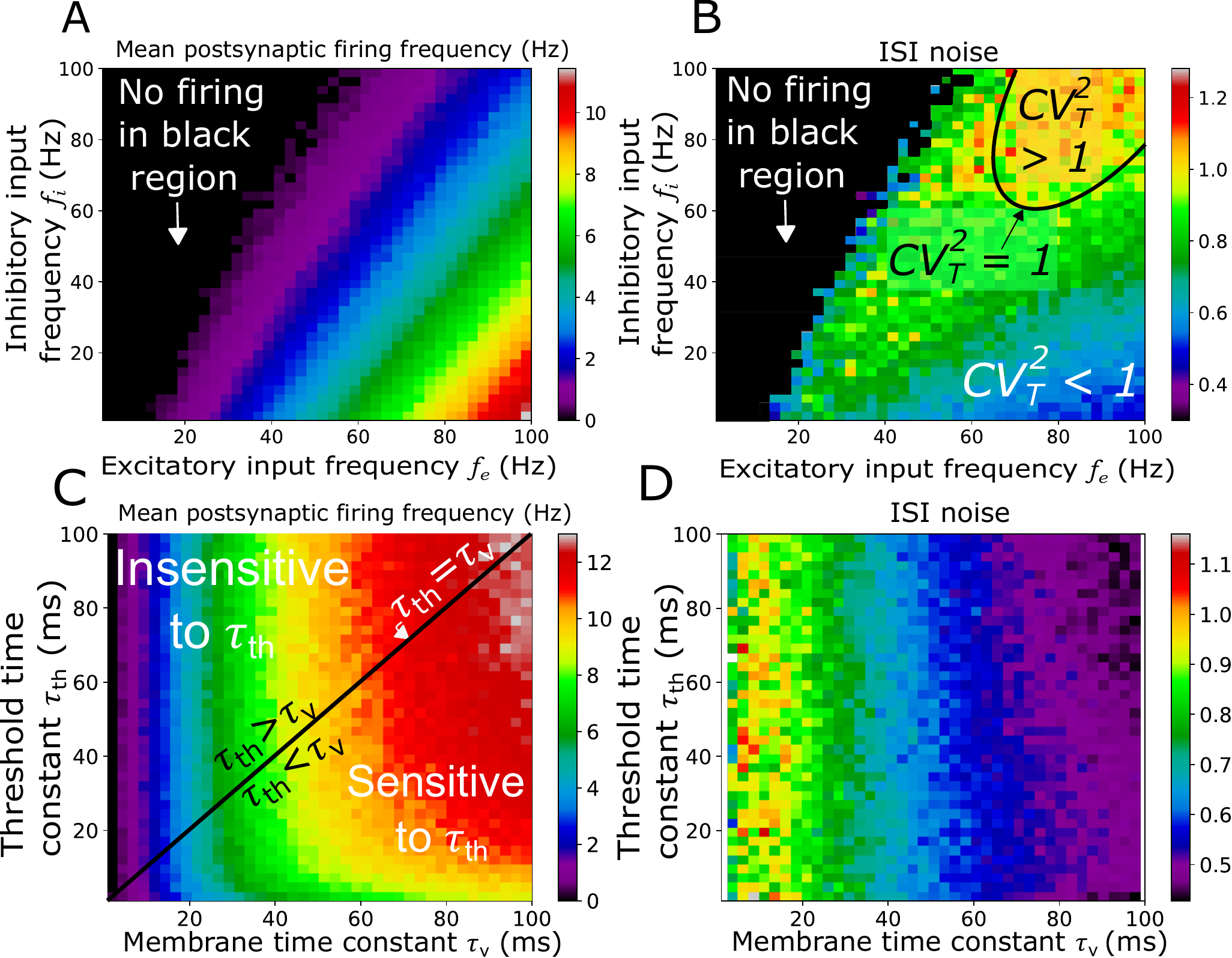}
\end{center}
\caption{\textbf{The mean postsynaptic firing frequency of a neuron with a depolarizing adaptive threshold increases with excitatory input frequency.
}  $\mathbf{A}$: The mean postsynaptic firing frequency as a function of the excitatory and inhibitory input frequency. $\mathbf{B}$:  The ISI noise as a function of the excitatory and inhibitory input frequency.  The curved black line represents the excitatory-inhibitory input frequency space where noise, defined by $CV_T^2$, is $1$.  The space contained within the curved line has $CV_T^2>1$ and $CV_T^2<1$ outside of this curved line. $\mathbf{C}$:  The mean postsynaptic firing frequency as a function of the threshold and membrane time constants.  The black line shows the mean postsynaptic firing frequency for $\tau_{th}=\tau_v$.  Above and below the black line shows the mean postsynaptic firing frequency for $\tau_{th}>\tau_v$, and $\tau_{th}<\tau_v$, respectively.   $\mathbf{D}$:  The ISI noise as a function of the threshold and membrane time constants.  The excitatory and inhibitory QC for each of these plots is modeled as binomial distributions with means $\langle b_e \rangle  =1$ and noise $CV_{b_e}^2 = 0.95$.   Parameters: $\tau_v=100 \ ms$, $\tau_{th}=100 \ ms$, $v_b = 20 \ mV$, $v_{min} = -50 \ mV$, $v_I = 10 \ mV$, $a_T = 0.8 \ mV, \ c_e = 2 \ mV, \ v_r = 0 \ mV$.}
\label{fig:dat2}
\end{figure*}

\section{Excitatory-inhibitory EI neuronal network with a depolarizing adaptive threshold}

In the previous section, we formulated the model of an excitatory and an inhibitory presynaptic neuron independently firing APs onto a postsynaptic neuron, which we called the EI model.  This model was analyzed in prior works \cite{gambrell2026adaptive} with a fixed threshold, an assumption which ignores the random nature of the threshold potential \cite{VERGARA1998321}.  We now improve the model by allowing the threshold potential to adapt with presynaptic firing activity.  Inspired by models of sodium channel kinetics \cite{platkiewicz2011impact} and similar to \cite{Lubejko2019SpikeThreshold}, the threshold potential is now modified to be a function of the membrane potential.

Specifically, the threshold potential now evolves via
\begin{equation}
    \frac{d v_{th}}{dt} = \frac{v_{th,\infty}(v)-v_{th}}{\tau_{th}}, \label{dvth1}
\end{equation}

where $\tau_{th}$ is a positively-valued threshold time constant, and $v_{th,\infty}(v)$ is the steady-state threshold potential
\begin{equation}    
v_{th, \infty }(v) :=
\begin{cases} 
v_{b}, & v < v_I, \\[1ex]
v_{b}+a_T (v-v_I), & v \ge  v_I. 
\label{vthpos}
\end{cases}
\end{equation}
In equation \eqref{vthpos}, the parameter $v_I$, which is the membrane potential level where the $v_{th, \infty}$ begins to increase, is called the inflection potential. The parameter  $a_T$, which controls the rate of increase of $v_{th,\infty}$, is called the scaling parameter.  A plot of equation \eqref{vthpos} is shown in Fig.~\ref{fig:dat}-A, and Fig.~\ref{fig:dat}-B shows trajectories of the depolarizing adaptive threshold in \eqref{dvth1} and \eqref{vthpos} and the membrane potential.  

Fig.~\ref{fig:dat}-C plots the mean postsynaptic firing frequency as a function of $f_e$ for both the fixed and depolarizing adaptive threshold models.  This figure shows that the mean postsynaptic firing frequency decreases with an adaptive threshold potential, as the threshold increases with excitation.   

Fig~\ref{fig:dat}-D plots the ISI noise as a function of the excitatory input frequency for both the fixed and depolarizing adaptive threshold models.  For each value of $f_e$, the mean postsynaptic firing frequencies of the two models were matched by incrementally adjusting the adaptive threshold parameter $v_b$ until the difference between the mean postsynaptic firing frequencies was within a $10 \ \mu s$ difference. The ISI noise peaks at intermediate frequencies for both models, with a larger maximum in the adaptive case due to the variability introduced by the changing of both $v(t)$ and $v_{th}(t)$.  The results in Fig.~\ref{fig:dat} are consistent with experimental results showing that dynamic threshold mechanisms contribute substantially to spike-time variability \cite{Badel2008}.

Having analyzed the ISI statistics for excitation alone, we now include inhibition.  Fig.~\ref{fig:dat2}-A and Fig.~\ref{fig:dat2}-B show the mean postsynaptic firing frequency and ISI noise as a function of the excitatory and inhibitory input frequencies, respectively.  Fig.~\ref{fig:dat2}-A shows that the mean postsynaptic firing frequency decreases with increasing $f_i$.  Fig.~\ref{fig:dat2}-B plots the ISI noise as a function of $f_e$ and $f_i$.  This plot shows that the excitatory-inhibitory input frequency space is partitioned into regions where $CV_T^2<1$, $CV_T^2=1$, and $CV_T^2 > 1$.  Noise is maximized when $f_e \approx f_i$ and $f_e$ and $f_i$ are both large, corresponding to the case where $CV_T^2>1$.  ISI noise is maximized under these conditions as $v(t)$ increases from the arrival of excitatory presynaptic APs at the same rate that it decreases due to the arrival of inhibitory presynaptic APs.  The increase in ISI noise when excitation and inhibition are approximately balanced is consistent with experimental observations \cite{Softky1993} of highly irregular cortical firing.

For the fixed threshold model, the mean postsynaptic firing frequency and the ISI noise were analyzed for the case of $\tau_v\to\infty$ and $\tau_v\to 0$, and in previous works \cite{gambrell2026adaptive}, the case of $0 < \tau_v < \infty$ was analyzed.  As variability is now introduced by a changing threshold, we now extend this analysis to determine how both $\tau_v$ and $\tau_{th}$ affect the ISI statistics.  Fig.~\ref{fig:dat2}-C plots the mean postsynaptic firing frequency as a function of both the threshold and membrane time constant, and shows that the mean postsynaptic firing frequency is insensitive to changes in $\tau_{{th}}$ when $\tau_{{th}}>\tau_v$.  Specifically, for a given range of $\tau_v$ where $\tau_{th}>\tau_v$, the mean postsynaptic firing frequency is independent of $\tau_{th}$.  This occurs because a threshold potential with a large threshold time constant changes so slowly with the membrane potential that it effectively behaves like a fixed threshold.  For $\tau_{{th}}<\tau_v$, the mean postsynaptic firing frequency becomes sensitive to changes in $\tau_{{th}}$.  This occurs because the threshold potential changes quickly with the membrane potential when $\tau_{th}$ is small.  When $\tau_{th} \approx 0$, the threshold potential changes at the same rate as the membrane potential, which prevents postsynaptic firing as no threshold crossings can occur.  

Fig.~\ref{fig:dat2}-D plots the ISI noise as a function of the threshold and membrane potential, and shows that noise is largest for small $\tau_v$.  Additionally, as $\tau_v$ increases, the ISI noise begins to show sensitivity to changes in $\tau_{th}$.  This occurs because a larger membrane time constant results in $v(t)$ remaining between $v_I$ and $v_{th}$ for longer durations, resulting in a larger increase in $v_{th}$.  Having extended the model by allowing the threshold potential to increase in response to excitatory presynaptic APs, we now explore the case where the threshold potential polarizes in response to inhibitory presynaptic APs. 

\begin{figure*}
\begin{center}
\includegraphics[width=0.8\linewidth]{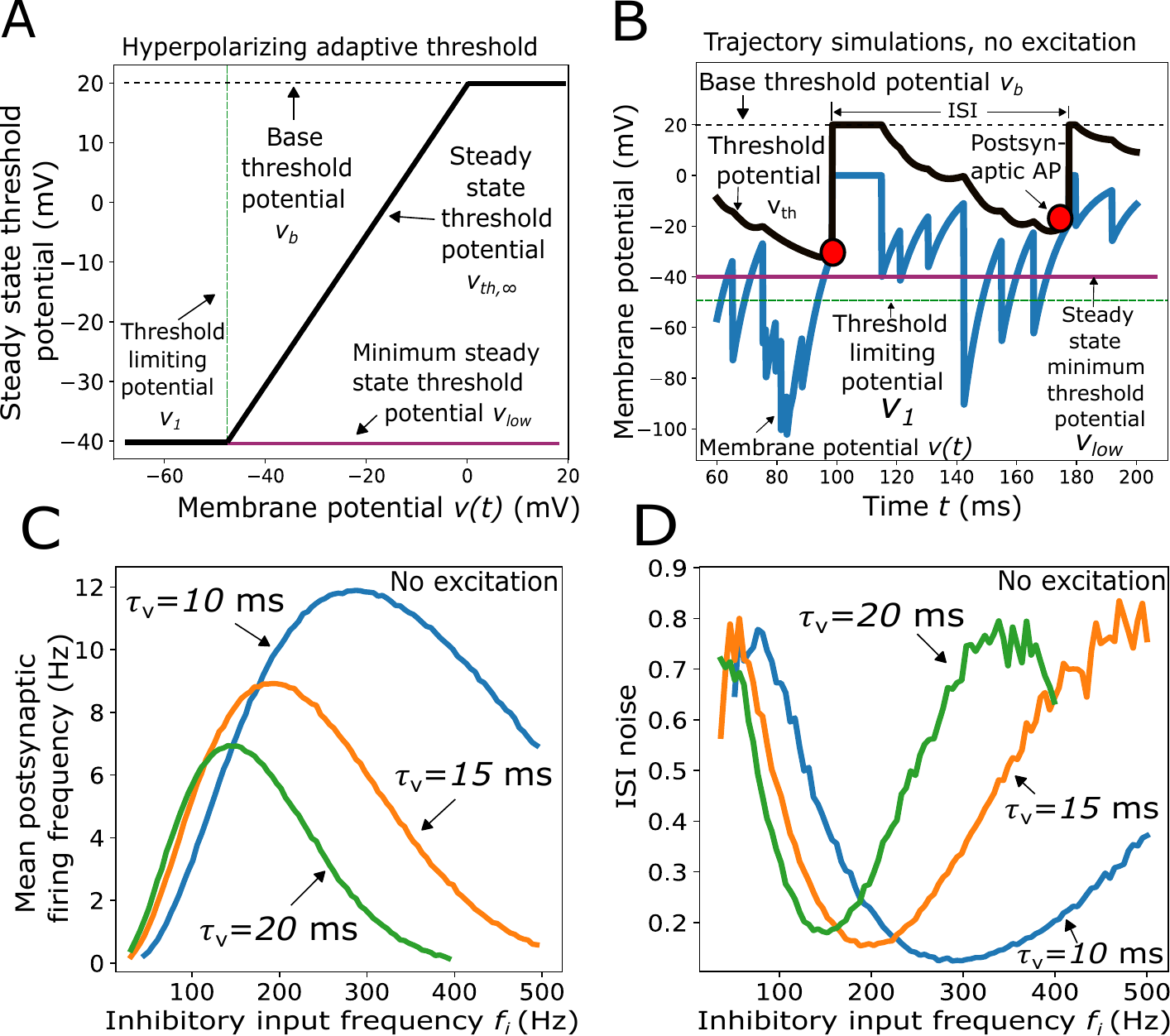}
\end{center}
\caption{\textbf{Presynaptic inhibition can elicit postsynaptic action potentials (APs) when the threshold potential decreases with a hyperpolarizing membrane potential.
}  $\mathbf{A}$:  The steady state threshold potential as a function of the membrane potential as in \eqref{vthinf2}.  $\mathbf{B}$: Trajectories of the hyperpolarizing adaptive threshold and membrane potential as a function of time.  The membrane potential (blue line) evolves via \eqref{vplus} and \eqref{dv}, with $f_e=0$, implying that the postsynaptic neuron receives only inhibition.  The threshold potential (black line) evolves via \eqref{dvth2} and \eqref{vthinf2}.  Once the membrane potential is driven to the threshold potential, the postsynaptic neuron fires an AP (red dot).  The inter-spike interval (ISI) is the time interval between successive postsynaptic APs. $\mathbf{C}$:  The mean postsynaptic firing frequency vs the inhibitory input frequency for $\tau_v = 10 \ ms$ (blue line), $15 \ ms$ (orange line), and $20 \ ms$ (green line). The mean postsynaptic firing frequency is maximized at intermediate inhibitory input frequencies, and the maximum value increases with decreasing membrane time constant.  $\mathbf{D}$:  The ISI noise as a function of the inhibitory input frequency for $\tau_v = 10 \ ms$ (blue line), $15 \ ms$ (orange line), and $20 \ ms$ (green line). In $\mathbf{A-D}$, the inhibitory QC is modeled as a binomial distribution with mean $\langle b_i \rangle=1$ and noise $CV_{b_i}^2 = 0.95$.   Parameters: $\tau_v=10 \ ms$, $\tau_{th}=20 \ ms$, $v_b = 20 \ mV$, $v_{1} = -50 \ mV$, $v_{low} = -40 \ mV$, $c_i= 2 \ mV, \ v_r = 0 \ mV$.}
\label{fig:hth}
\end{figure*}

\section{Excitatory-inhibitory neuronal EI network with a hyperpolarizing adaptive threshold}

In the previous section, we extended the EI model by allowing the threshold potential to increase with the membrane potential.  This increase models the probabilistic nature of the threshold potential in the presence of excitation.  In the absence of excitation, however, inhibitory presynaptic APs have been shown to induce postsynaptic APs \cite{MAHROUS20251906}.  We now allow the threshold potential to decrease with a hyperpolarizing membrane potential, modeling sodium channel \cite{Bean2007} recovery from inactivation.  We refer to this model as a hyperpolarizing adaptive threshold model.

In order to model the observation that inhibition can induce postsynaptic APs, we allow the threshold potential to depend on the membrane potential and evolve via 
\begin{equation}
    \frac{d v_{th}}{dt} = \frac{v_{th,\infty}(v)-v_{th}}{\tau_{th}}, \label{dvth2}
\end{equation}
as before, where $\tau_{th}$ is the threshold time constant that controls the rate at which the threshold potential approaches a steady-state level $v_{th, \infty}(v)$.  We construct the steady-state threshold level such that the threshold potential behaves as follows: in the case where the membrane potential is at rest ($v=v_r$), the threshold potential returns to a base value $v_b$ at steady state. But for the hyperpolarized case ($v<v_r$), the threshold potential decreases with decreasing $v(t)$ to a steady state low level $v_{low}$ which occurs when $v=v_1$.  We refer to $v_1$ as the threshold limiting potential.  Furthermore, it is assumed that $v_1<v_{low}$, as otherwise postsynaptic APs are always capable of occurring once $v<v_1$.  We modeled the steady state threshold potential described as
\begin{equation}
v_{th,\infty}(v) :=
\begin{cases} 
v_{low}, & v < v_1, \\[1ex]
v_{low} - \dfrac{v - v_1}{v_1} \,(v_{b} - v_{low}), & v_1 \le v < v_r, \\[1ex]
v_{b}, & v=v_r.\label{vthinf2}
\end{cases}
\end{equation}
Fig.~\ref{fig:hth}-A shows a plot of \eqref{vthinf2}, and Fig.~\ref{fig:hth}-B shows trajectories of the hyperpolarizing threshold potential in \eqref{dvth2} and the membrane potential in \eqref{vplus} and \eqref{dv} as a function of time. 

In Fig.~\ref{fig:hth}-C, the mean postsynaptic firing frequency is shown as a function of the inhibitory input frequency for three different levels of $\tau_v$. This figure shows that the mean postsynaptic firing frequency is maximized at intermediate inhibitory input frequencies. This occurs for two reasons. First, for sufficiently low $f_i$, the threshold potential never drops below the resting potential, and no threshold crossings can occur. Second, for sufficiently large $f_i$, the membrane potential drops below the steady state minimum threshold potential $v_{low}$, again preventing threshold crossings. Furthermore, the maximum mean postsynaptic frequency decreases as $\tau_v$ increases. This occurs because a larger $\tau_v$ causes $v(t)$ to return more slowly to $v_r$, allowing $v_{th}$ more time to return to $v_b$ and reducing the number of threshold crossings.

Fig.~\ref{fig:hth}-D shows ISI noise, defined by $CV_T^2$, as a function of $f_i$ for the same membrane time constants in Fig.~\ref{fig:hth}-C.  From this figure, we can see that noise is minimized at intermediate $f_i$, and the value of $f_i$ where this minimum occurs increases with increasing $\tau_v$.  This occurs because the membrane potential returns more quickly to the resting potential with a larger time constant, which increases the chance of a threshold crossing.   

\begin{figure*}
\begin{center}
\includegraphics[width=0.7\linewidth]{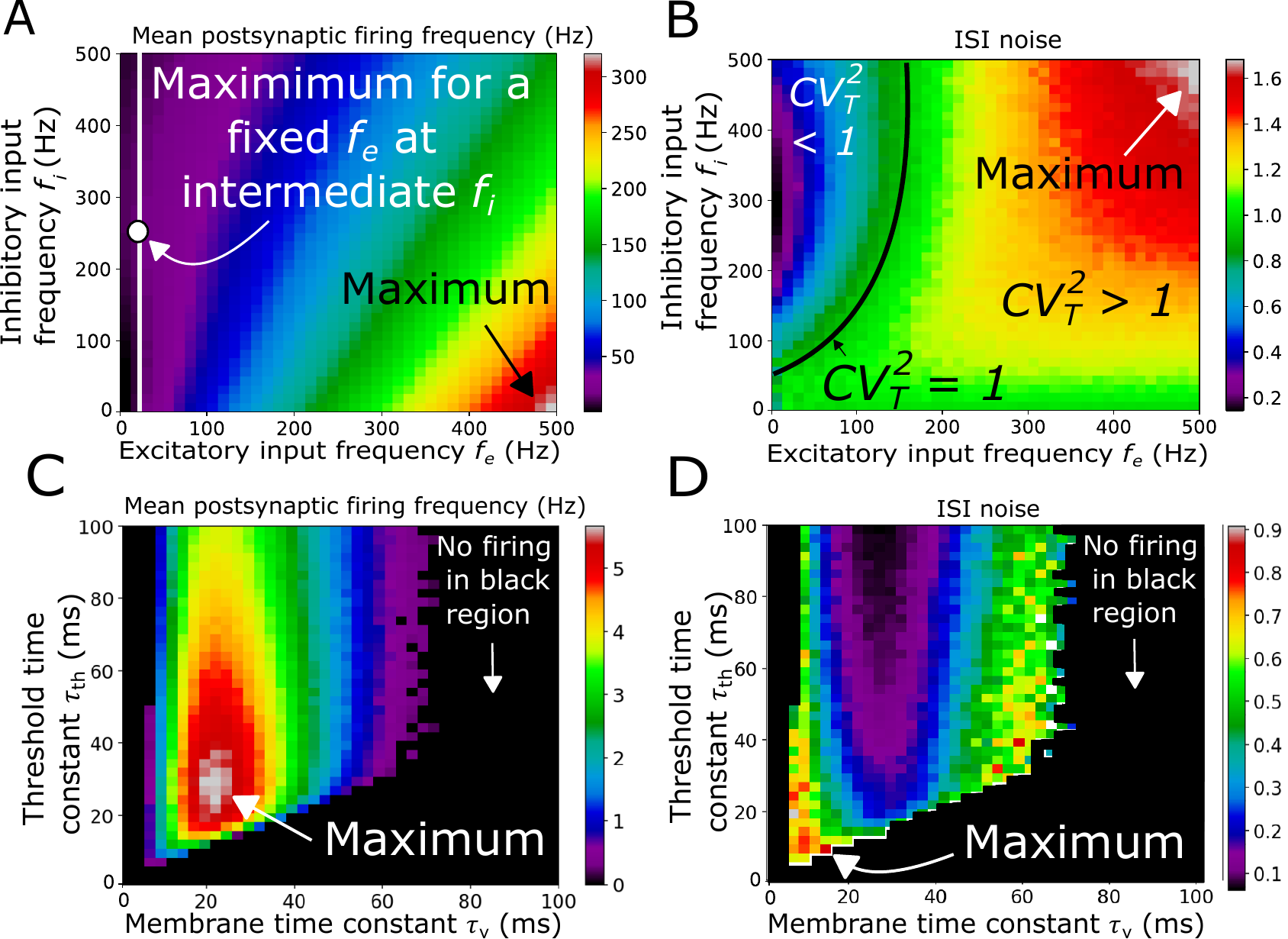}
\end{center}
\caption{\textbf{The mean postsynaptic firing rate of a neuron with a hyperpolarizing adaptive threshold is maximized at intermediate membrane and threshold time constants.
}  $\mathbf{A}$: The mean postsynaptic firing frequency as a function of the excitatory and inhibitory input frequency.  The white line shows the mean postsynaptic firing frequency for $f_e=20 \ Hz$, and the white circle on the white line represents the maximum postsynaptic firing frequency occurring at $f_i=265 \ Hz$. $ \mathbf{B}$:  The ISI noise as a function of the excitatory and inhibitory input frequency.  The black line represents the parameter space of $f_e$ and $f_i$ where $CV_T^2=1$.  In the space to the left of the black line $CV_T^2<1$, and to the right of the black line $CV_T^2>1$. $\mathbf{C}$:  The mean postsynaptic firing frequency as a function of the threshold and membrane time constant.  The mean postsynaptic firing frequency is maximized at intermediate values of the time constants.  $\mathbf{D}$:  The ISI noise as a function of the threshold and membrane time constants.  In $\mathbf{A}$-$\mathbf{D}$, the excitatory and inhibitory QC are modeled as a binomial distribution with means $\langle b_e \rangle  =\langle b_i \rangle=1$ and noise $CV_{b_e}^2=CV_{b_i}^2 = 0.95$.   Parameters: $\tau_v=10 \ ms$, $\tau_{th}=20 \ ms$, $v_b = 20 \ mV$, $v_{1} = -50 \ mV$, $v_{low} = -40 \ mV$, $c_i= 2 \ mV, \ c_e = 2 \ mV, \ v_r = 0 \ mV$.}
\label{fig:hth2}
\end{figure*}

Having analyzed the ISI statistics in terms of inhibition only, we now explore how the addition of excitation affects the ISI statistics.  Fig.~\ref{fig:hth2}-A plots the mean postsynaptic firing frequency as a function of both the excitatory and inhibitory input frequency.  This figure shows that the mean postsynaptic firing frequency is maximized for low $f_i$ and high $f_e$.  Additionally, at low values of $f_e$, the mean postsynaptic firing frequency is maximized at intermediate $f_i$, as shown in Fig.~\ref{fig:hth}-C. Fig.~\ref{fig:hth2}-B plots the ISI noise as a function of $f_e$ and $f_i$, and shows three things.  First, for $f_i<80 \ Hz$, the ISI noise is approximately fixed.  Second, when $f_e\approx 150 \ Hz$, the ISI noise reaches a minimum value at intermediate $f_i$.  And third, for $f_e>150 \ Hz$, the ISI noise increases with increasing $f_i$.

Having investigated how $f_e$ and $f_i$ affect the ISI statistics, we now study how $\tau_v$ and $\tau_{th}$ affect the ISI statistics, as they are important components of the model that govern the timing of threshold crossings. In Fig.~\ref{fig:hth2}-C, the mean postsynaptic firing frequency is plotted as a function of the threshold and membrane time constant.  Interestingly, the maximum mean postsynaptic firing frequency occurs when both time constants are low and the threshold time constant is slightly larger.  This occurs for two reasons. The first is that a large $\tau_{th}$ slows the rate at which the threshold potential decreases, which leads to longer ISIs.  When $\tau_{th}$ is sufficiently large, $v_{th}$ changes so slowly that it behaves like a fixed threshold, and no postsynaptic APs can occur. Second, when $\tau_{th}$ is sufficiently small, the threshold potential changes too quickly for threshold crossings to occur.  Additionally, if the membrane time constant is too small, the threshold potential remains near its base threshold potential, and no threshold crossings occur. As the threshold time constant becomes too large, the membrane potential decreases without bound, preventing any crossings from occurring.

Fig.~\ref{fig:hth2}-D shows ISI noise, defined by $CV_T^2$, as a function of the threshold and membrane time constants.  This figure shows that the ISI noise increases for a given $\tau _ {v} $ as $\tau_{th}$ increases. This occurs because the threshold potential changes more slowly, leading to an increased number of threshold crossings at lower membrane time constants.  Additionally, the ISI noise is maximized at low values of $\tau_v$ and $\tau_{th}$.  This maximum randomness at low time constants is due to the fact that both $v(t)$ and $v_{th}$ are changing quickly, which introduces variability into the threshold crossing times.

\section{Conclusion}

This paper analyzes the statistics of the inter-spike interval (ISI) for a postsynaptic neuron that independently receives excitatory and inhibitory presynaptic APs (see Fig.~\ref{fig:schemtraj}) for three types of thresholds:  a fixed threshold, a threshold potential that increases in response to presynaptic excitation, and a threshold potential that decreases in response to presynaptic inhibition.  We refer to this model as the EI model.   For the EI model with a fixed threshold, an analytical equation for the mean ISI for the case of $\tau_v\to\infty$ was derived (see Fig.~\ref{fig:ft}-A).  Following this, the ISI statistics were discussed for other parameter spaces for a fixed threshold.

Next, the threshold potential was allowed to increase in response to presynaptic excitation (see Fig.~\ref{fig:ft}), and we analyzed the mean postsynaptic firing frequency and ISI noise as a function of the excitatory and inhibitory input frequencies and the threshold and membrane time constants.  Our results showed that the ISI statistics as a function of the excitatory and inhibitory input frequencies were similar in form to the results of the fixed threshold model.  Furthermore, when we analyzed the ISI statistics as a function of the threshold and membrane time constants, our results showed that the mean postsynaptic firing frequency was insensitive to changes in the threshold time constant when it exceeded the membrane time constant (see Fig.~\ref{fig:dat}, Fig.~\ref{fig:dat2}).  Additionally, the maximum mean firing frequency was shown to increase as both the threshold and membrane time constants increased. ISI noise was shown to exhibit an increased sensitivity to changes in the threshold time constant with increasing membrane time constant.

These findings are consistent with experimental observations demonstrating that neuronal spike threshold is dynamically modulated by membrane potential and recent activity rather than remaining fixed. In vivo recordings \cite{Azouz2000} have shown that threshold variability can substantially influence spike timing and coincidence detection. Our results extend this idea by showing that adaptive threshold dynamics can strongly shape inter-spike interval statistics, increasing or decreasing firing variability depending on the form of threshold adaptation.

After this, we allowed the threshold potential to decrease in response to presynaptic inhibition, modeling postinhibitory facilitation (see Fig.~\ref{fig:hth}-A).  Our results showed that the mean postsynaptic firing frequency is maximized at intermediate inhibitory input frequencies.  Subsequently, the mean postsynaptic firing frequency and its noise were studied as a function of the excitatory and inhibitory input frequencies.  Our analysis showed that the mean postsynaptic firing frequency was maximized at intermediate inhibitory input frequencies at low excitatory input frequencies.  Additionally, the ISI noise at high inhibitory input frequencies was shown to either reach a local minimum value or a maximum value, depending on the excitatory input frequency level.  In contrast to the case where the threshold potential increased with excitation, the mean postsynaptic firing frequency was shown to be maximized at intermediate values of the threshold and membrane time constant (see Fig.~\ref{fig:hth2}-C).  ISI noise was shown to decrease with increasing threshold time constant.

In future works, the first-passage-time framework will be utilized to derive equations for the ISI statistics for both the fixed and adaptive threshold models, improving the predictive power of the models.  Furthermore, the threshold potential in both adaptive models will be changed to nonlinear functions, modeling the nonlinear effects of ion channels on the threshold potential \cite{Brette2005}.  Finally, the model will be extended to study feed-forward and feedback networks.



\end{document}